\documentclass[a4paper,fleqn,usenatbib]{mnras}

\usepackage[T1]{fontenc}
\usepackage{ae,aecompl}
\usepackage{graphicx}   
\usepackage{amsmath}    
\usepackage{amssymb}    

\newcommand{\source}{\hbox{3C\,277.3}}
\newcommand{\pks}{\hbox{PKS\,B2152-699}}

\newcommand{\chandra}{\textit{Chandra}}

\newcommand{\hst}{\textit{HST}}
\newcommand{\vla}{\textit{VLA}}
\newcommand{\frI}{\hbox{FR\,I}}
\newcommand{\fr}{\hbox{FR\,I/II}}
\newcommand{\frII}{\hbox{FR\,II}}

\title[The X-ray emission of \source (Coma A)]
{X-rays associated with the jet-cloud interacting radio
  galaxy \source\ (Coma A): implications for energy deposition}

\author[D.M. Worrall et al.]
 {D.M. Worrall$^{1,2}$,
M. Birkinshaw$^{1,2}$
and A.J.~Young$^1$
\\
$^1$HH Wills Physics Laboratory, University of Bristol, Tyndall
Avenue, Bristol BS8~1TL \\
$^2$Harvard-Smithsonian Center for Astrophysics, 
60 Garden Street, 
Cambridge, MA\ \ 02138, USA \\
}

\begin{document}

\label{firstpage}

\maketitle

\begin{abstract}

We report the discovery with \chandra\ of X-ray-emitting gas
associated with the jet-cloud interaction in the radio galaxy
\source\ (Coma A), a source that falls in the most important power
range for radio-mode feedback in the Universe.  This hot gas, heated
by the jet, dominates the mass of the cloud which is responsible for
an extreme projected deflection of the kpc-scale radio jet. 
Highly absorbed X-ray emission from the nucleus of
\source\ confirms that the jet lies close to the plane of the sky and
so has a large intrinsic deflection.  We detect group gas on the scale
of the radio lobes, and see X-ray cavities coincident with the
brightest radio emission, with the lobes embraced by X-ray
enhancements that we argue are the result of shocks. The
anti-correlation between the locations of X-ray arms and
H$\alpha$-emitting filaments that are believed to have originated from
a merger with one or more gas-rich galaxies suggests that shocks
advancing around the lobe are inhibited by the dense colder material.
Synchrotron X-ray emission is detected from the upstream edge of a
second bright radio knot.  X-rays are also detected from the location
where an undetected counterjet enters the northern radio hotspot.  We
suggest that these X-rays are synchrotron radiation from a shock in a
small-scale substructure.

\end{abstract}


\section{Introduction}
\label{sec:intro}
The persistence of radio galaxies means that the sources must be
expanding into an external medium.  X-ray-emitting plasma held in the
potential wells of galaxies, groups, and clusters dominates that
medium.  Dynamical studies are hampered both by the fact that
radio-source emission is predominantly non-thermal, and by
instrumentation that cannot yet provide the spatially-resolved
high-resolution X-ray spectroscopy needed to probe the dynamics of the
X-ray gas through its line emission.  If the X-ray gas is mixed with
cooler ionized material that emits lines in the optical or near
infrared, or neutral gas emitting in the radio or sub-mm, then
dynamical studies are greatly enhanced.  For this reason, the study of
radio galaxies with cooler emission-line gas in their environments is
of particular interest, especially those displaying jet-cloud
interactions, where an interaction interface is directly revealed.  A
prominent case previously unstudied in X-rays is \source\ (Coma A).

There is an additional reason that \source\ is important: its radio
power and morphology place it on the borderline between the \frI\ and
\frII\ classes.  Radio galaxies are recognized as the primary sources
for the energy needed to maintain cluster cores, and for playing a
dominant r\^ole in the self-regulated feedback loop between black-hole
and galaxy growth \citep{mcnamara-nulsen}.  An understanding of how
radio-mode feedback actually works remains incomplete, and is based
primarily on observations of radio sources that are nearest and most
abundant, and hence members of the \frI\ population.  The scaling of
cavity and radio power that is found \citep{cavagnolo}, when combined
with the radio luminosity function \citep{best}, means that radio-mode
feedback is dominated by sources of \fr\ transition power
\citep{wrev}, with half the total heating expected from sources with
powers $0.3 - 3$ times the \fr\ transition power.  Interestingly the
jet-cloud interacting radio source \pks\ is also of \fr\ transition
power.  \pks\ has been the subject of in-depth study with
\chandra\ \citep*{ly, young, wfos}, and makes an interesting
comparison case for \source.

3C\,277.3 is a relatively small double-lobed radio galaxy (total size
about 65 kpc), studied in detail by \citet{vanbreugel}.  In the
northern lobe there is a prominent hotspot but no obvious jet, whereas
in the southern lobe two radio knots delineate a dramatic bend in the jet
through a projected angle of about $40^\circ$, but there is no
hotspot.  While also known as Coma A, \source\ lies behind the Coma
cluster and is offset from its centre by about 75 arcmin,
outside the boundary of cluster X-ray emission.

The entire \source\ system is rich in cool gas emitting in a number of
optical lines \citep{vanbreugel} and cold hydrogen is seen in
absorption at 21 cm \citep{morganti}.  The H$\alpha$ emission-line gas
forms a system of large-scale arcs and filaments which extend almost
as far perpendicular as parallel to the radio axis, and which appear
to bound the radio lobes \citep{tadhunter}. 
What is exciting this gas is uncertain, and while \citet{tadhunter}
favour shocks as being dominant, they do not rule out
photoionization from young stars in the filaments.
The observation of
resolved H\,I absorption that is detected against both radio lobes
with velocity structure matched to that seen in emission-line gas has
led \citet{morganti} to suggest that the ionized and neutral gas are
two phases of the same structure, of total gas mass at least $10^9$
M$_\odot$, resulting from a merger of \source's host with at least one
large gas-rich galaxy.  The cold gas is deduced to be filamentary on a
sub-arcsec scale: an anti-correlation of radio polarization and
H$\alpha$ emission, particularly in the northern hotspot and around
the southern jet, is interpreted as the result of differential Faraday
rotation from multiple clumps of thermal material lying within the
radio beam \citep{vanbreugel, baum}.  The general conclusion is that
\source\ is expanding into a gas disk \citep{morganti}.

3C\,277.3's designation as a jet-cloud interacting radio galaxy is due
to the bright optical emission-line knot that lies adjacent to the
large bend in the southern jet, and with which the jet is interacting
\citep{vanbreugel, si, tilak}.

\citet{balmaverde} comment on blobs of X-ray emission cospatial with
two radio knots and the northern hotspot.  However, the quality of the
data was insufficient to locate the X-ray features accurately, address
emission mechanisms, or search for faint diffuse emission.  Here we
discuss much improved \chandra\ X-ray data, revealing that the
jet-cloud interaction region is dominated in mass by X-ray emitting
gas, and that the system resides in a group atmosphere.  We estimate
the level of power injected by the radio source into the cloud and
into the larger-scale group gas, thus addressing the heating of the
external medium by a characteristic radio galaxy in a characteristic
atmosphere, and so important for the Universe as a whole.  
The source is classified by its nuclear spectrum as a High
  Excitation Galaxy \citep[HEG;][]{butt1, butt2}, and we also
report here on the X-ray emission from the nucleus, as well as
that from the southern jet and northern
hotspot.  We adopt values for the cosmological parameters of $H_0 =
70$~km s$^{-1}$ Mpc$^{-1}$, $\Omega_{\rm {m0}} = 0.3$, and
$\Omega_{\Lambda 0} = 0.7$.  Thus 1~arcsec corresponds to a projected
distance of 1.6~kpc at \source, for its redshift of $z =0.0853$
\citep{rines}.

\section{Observations and reduction methods}
\label{sec:obs}%

\subsection{\chandra\ X-ray}
\label{sec:xobs}

We observed \source\ in VFAINT full-frame data mode with the Advanced
CCD Imaging Spectrometer (ACIS) on board \chandra\ in four exposures
taken between 2014 March 11th and 16th (OBSIDs 15023, 15024, 16599,
16600).  The source was positioned at the nominal aimpoint of the
front-illuminated I3 chip. Our analysis combines these data with those
from an archival ACIS exposure \citep[OBSID 11391,][]{balmaverde}
taken on 2010 March 3rd in the same data mode but with the source
positioned at the nominal aimpoint of the back-illuminated S3 chip.
Details of the ACIS instrument and its modes of operation can be found
in the \chandra\ Proposers' Observatory Guide\footnote{
  http://cxc.harvard.edu/proposer}.  Results presented here use {\sc
  ciao v4.7} and the {\sc caldb v4.6.8} calibration database.  We
re-calibrated the data, with random pixelization removed, the
energy-dependent sub-pixel event repositioning algorithm applied, and
bad pixels masked, following the software `threads' from the
\chandra\ X-ray Center (CXC)\footnote{ http://cxc.harvard.edu/ciao},
to make new level~2 event files.  Only events with grades 0,2,3,4,6
were used, as recommended.

The observations were free from background flares and, after removal
of time intervals when the background deviated more than $3\sigma$
from the average value, the total exposure time was 
213 ks, distributed as follows: OBSID 15023, 43.114 ks; 15024, 19.828
ks; 16599, 28.513 ks; 16600, 96.744 ks; 11391, 24.770 ks.  The
astrometry was adjusted (by between 0.5 and 1 arcsec, mostly in RA) so
that for each observation the X-ray core aligned with the radio-core
position measured from our highest-resolution 4.9~GHz map (Section
\ref{sec:robs}) of RA$=12^{\rm h} 54^{\rm m} 11^{\rm s}\llap{.}993$,
Dec$=27^\circ 37' 33''\llap{.}87$.

The {\sc ciao wavdetect} task was used to find sources in a merged
0.4--5-keV image of the 2014 data, with the threshold set at 1
spurious source per field.  Except for those detections corresponding
to features associated specifically with features of \source, the
corresponding regions were excluded from spectral analysis, and were
refilled using {\sc dmfilth} before making image figures.  All
spectral fitting includes absorption along the line of sight in our
Galaxy assuming a column density of $N_{\rm H} = 0.72 \times 10^{20}$
cm$^{-2}$ [from the {\sc colden} program provided by the CXC, using
data of \citet{dlock90}], including where models are described as unabsorbed.

Throughout the paper the power-law spectral index, $\alpha$, is defined in the sense that flux density is
proportional to $\nu^{-\alpha}$.  
X-ray spectral indices are quoted in
terms of $\alpha$ rather than the values one larger returned by
spectral-fitting codes.
Uncertainties correspond to 90 per
cent confidence, unless otherwise stated.

\subsection{\vla\ radio}
\label{sec:robs}

\begin{figure}
\centering
\includegraphics[width=6.5cm]{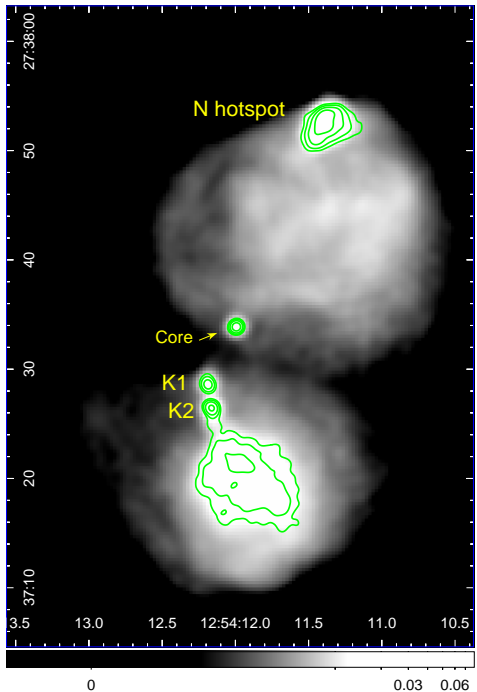}
\caption{ Radio structure of \source.  Image of 1.4-GHz data with
  1.3-arcsec beam (entry 3 of Table~\ref{tab:radio}) with scale bar in
  Jy, with contours from the 4.8-GHz data with 0.4-arcsec beam (entry
  1 of Table~\ref{tab:radio}) increasing by factors of $2$ from the
  lowest contour at 0.2 mJy beam$^{-1}$.}
\label{fig:radio}
\end{figure}

For this work we mapped \vla\ data from the programmes, arrays and
frequencies given in Table~\ref{tab:radio}.  The data were calibrated,
flagged for interference, and passed through the normal clean and gain
self-calibration cycles in {\sc aips}.  Figure \ref{fig:radio} shows
the overall structure and prominent components of the radio source.
The source has well-rounded diffuse lobes, a prominent northern (N)
hotspot, and two bright southern knots which are labelled K1 and K2,
following \citet{bridle}.

\begin{table*}
\caption{\vla/ Radio Data}
\label{tab:radio}
\begin{tabular}{llclcl}
\hline
(1) & (2) & (3) & (4) & (5) & (6) \\
Programme & Observation Date & Frequency
  (GHz) & Configuration & Restoring Beam
  (arcsec)$^2$ & Noise ($\mu$Jy beam$^{-1}$) \\
\hline
VANB$^a$ & 1981 Mar 18 & 4.87 &
\vla\ A & $0.40 \times 0.37$ & 78 \\
VANB$^a$ & 1981 May 31 & 4.89 &
\vla\ B & $1.28 \times 1.22$ & 54 \\
VANB$^a$ & 1981 Mar 18 & 1.41 &
\vla\ A & $1.38 \times 1.34$ & 133 \\
AB348 & 1985 Aug 26 & 14.94 &
\vla\ C & $1.53 \times 1.24$ & 111 \\
\hline
\end{tabular}
\medskip
\begin{minipage}{\linewidth}
$a$. Data originally published in \citet{vanbreugel}
\end{minipage}
\end{table*}

\section{Results}
\label{sec:results}

\subsection{Overall X-ray characteristics}
\label{sec:overall}%

\begin{figure}
\centering
\includegraphics[width=8.5cm]{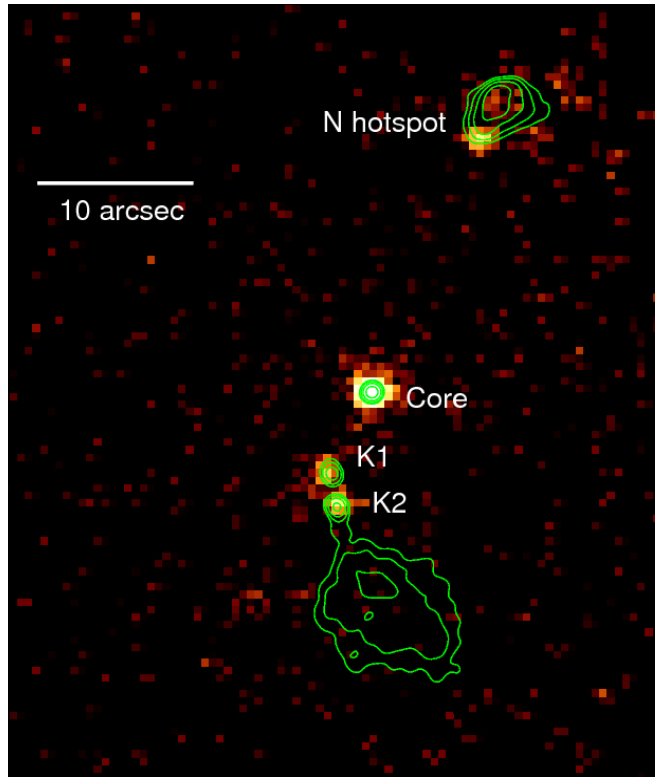}
\caption{Exposure-corrected 0.4--5-keV \chandra\ image in 0.492 arcsec
  pixels. Radio contours are from Fig.~\ref{fig:radio}.}
\label{fig:xrimage}
\end{figure}

The five \chandra\ data sets were exposure corrected and combined to
make the 0.4--5-keV combined image shown in Figure~\ref{fig:xrimage}.
Here X-rays associated with the core, two knots in the southern jet
(K1, K2) and the N hotspot are readily apparent.  Most pixels contain
no counts.

\begin{figure}
\centering
\includegraphics[width=8.0cm]{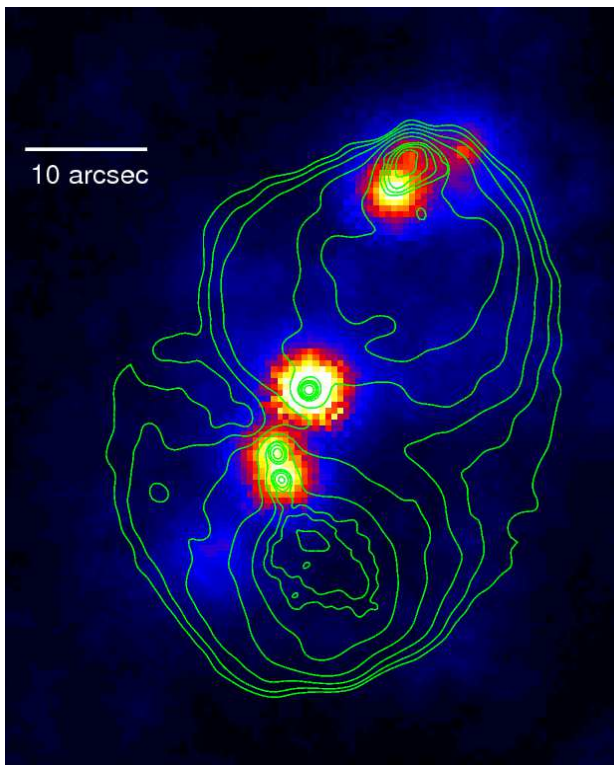}
\caption{Adaptively-smoothed exposure-corrected 0.4--5 keV
  \chandra\ image together with both the radio contours of
  Fig.~\ref{fig:radio} and those from the 1.4-GHz data with 1.3-arcsec
  beam (entry 3 of Table~\ref{tab:radio}) increasing by factors of $2$
  from the lowest contour at 0.4 mJy beam$^{-1}$.}
\label{fig:xradaptimage}
\end{figure}

The application of adaptive smoothing to the image in
Figure~\ref{fig:xrimage} shows underlying diffuse structure
(Fig.~\ref{fig:xradaptimage}), whereas the background should be
roughly uniform over the image.  The X-rays show a rough morphological
anti-correlation with the radio, with the X-rays appearing to embrace
the radio lobes.  Figure~\ref{fig:xradaptimage} shows X-ray
cavities in the brightest parts of the S and N lobes (disregarding the
N hotspot and S jet).  The region directly to the SW of the core is
faint in X-rays where the radio too is weak, but there is an apparent
X-ray increase further out, bordering the faintest radio contour in
the image.  The X-ray structures are reminiscent of the arms seen in
H$\alpha$ emission \citep{tadhunter}, a point to which we will return
in Section \ref{sec:gas}.

\begin{figure}
\centering
\includegraphics[width=5.0cm]{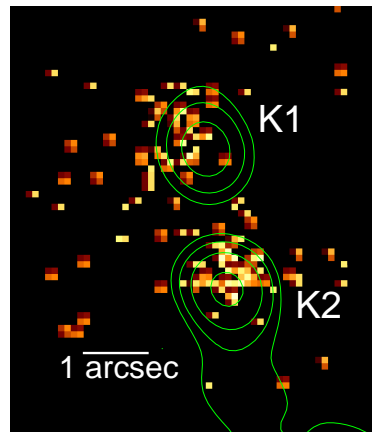}
\caption{Same as Figure~\ref{fig:xrimage} but zoomed to show the
  K1--K2 region in more detail, and with the X-ray mapped into sub
  pixels of 1/5th the native size (i.e., 0.0984 arcsec). Relative to
  the radio, the X-ray emission is displaced on the outer edge of the
  bend at K1 and upstream where the jet flows into K2.}
\label{fig:xrzoomK}
\end{figure}

The alignment of the N hotspot, core and K1 in the radio (e.g.,
Fig~\ref{fig:radio}) suggests that the southern radio jet enters K1
from the NW and deflects by a projected angle of almost 40 degrees to
reach K2.  The X-ray and radio data are registered on the core to
better than 0.1 arcsec, rendering the offsets between the radio and
X-ray in K1 and K2 (Fig.~\ref{fig:xrzoomK}) striking.  At K1 the X-rays
are displaced at the outer edge of the radio bend, and better
aligned with the optical emission-line gas discussed in Section
\ref{sec:hic}.  At K2 the X-rays are along the path of the radio jet,
although peaked upstream of the radio knot.

\begin{figure}
\centering
\includegraphics[width=6.5cm]{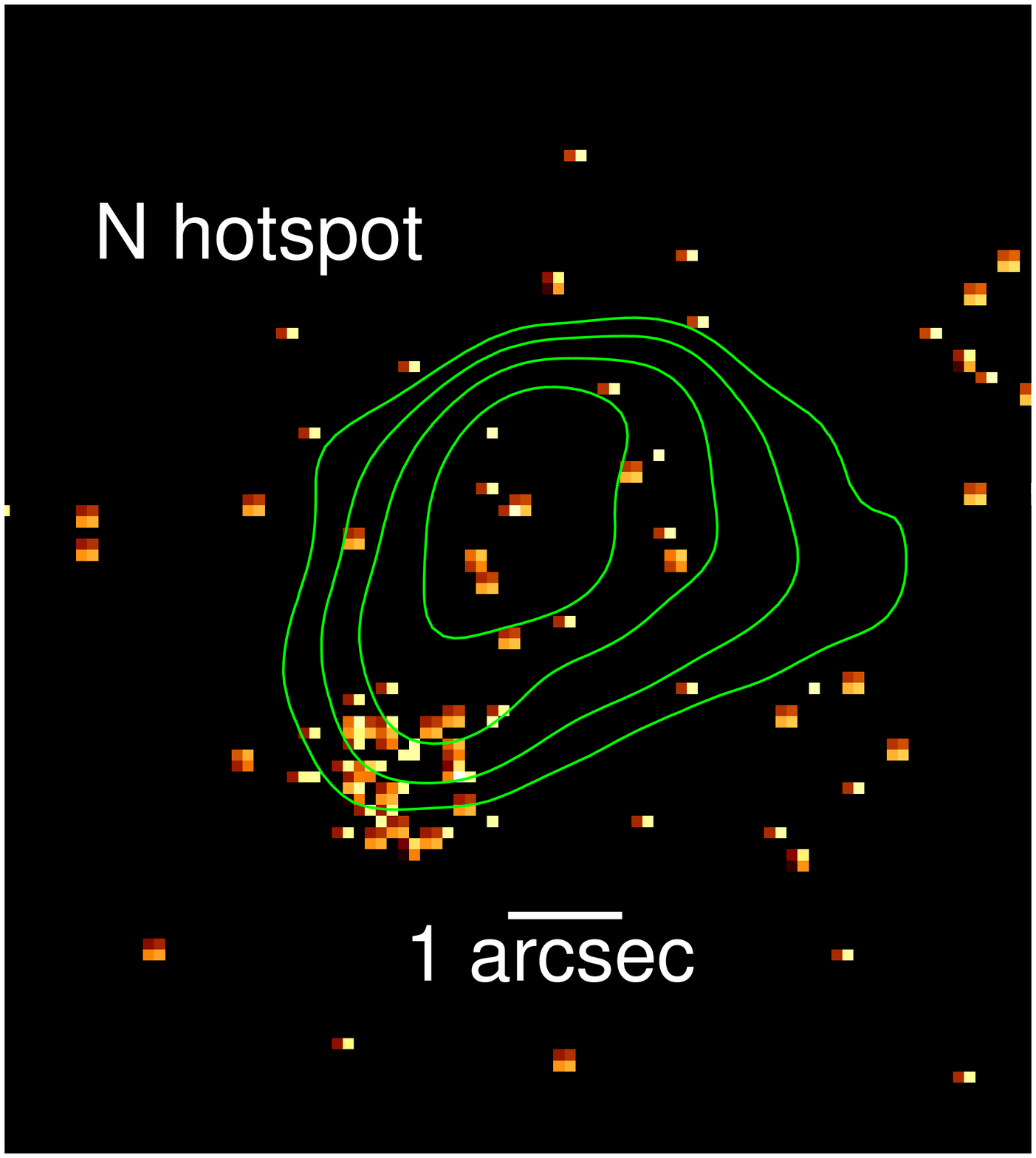}
\caption{Same as Figure~\ref{fig:xrzoomK} but for the N hotspot,
  showing that the X-rays are bright at the upstream edge, where the
  jet presumably enters.}
\label{fig:xrzoomNhot}
\end{figure}

There is X-ray emission covering the N hotspot, as clear from
Figures~\ref{fig:xrimage} and \ref{fig:xradaptimage}.  The brightest
X-ray emission is at the upstream edge, where the counterjet is
presumed to enter the hotspot (Fig.~\ref{fig:xrzoomNhot}).  While we
cannot rule out a foreground or background source as responsible for
these X-rays, our catalogue and image searches at other wavelengths,
including the use of archival \hst\ data, revealed no strong
candidates for an identification.  We note that the western X-ray
extension of the N hotspot seen in Figure~\ref{fig:xradaptimage} is
unreliable because here the exposure has dropped by a factor of two
due to proximity to a CCD chip gap, allowing the exposure correction
to emphasize a statistically small number of counts.

\subsection{Core}
\label{sec:core}%

We have fitted the spectrum (roughly 1200 net counts 0.4-7 keV)
extracted from a circle of radius 1.25 arcsec centered on the core,
using local background, to various models. The data and calibrations
for the 2014 observations were combined before fitting. Data were
grouped to 15 counts/bin and the weighting scheme of \citet{churazov}
was adopted to provide an improved estimate of the variance in the
limit of small numbers of counts.  While a single-component power-law
model gave a bad fit, it was noticeable that the spectral index was
abnormally hard, $\Gamma \approx 0$ (i.e., flux-density rising with
energy: $\alpha_{\rm x} \approx -1$), and there was structure in the
residuals, with a broad deficiency of counts between about 2 and 3 keV
and excess counts at high and low energies.  This suggested the
presence of an additional component that is heavily absorbed.  
  The distribution of spectral counts did not agree with that
normal for nearby \frI\ radio galaxies: a weakly absorbed power law, 
possibly mixed with a small component of
  thermal emission from central gas
  \citep[e.g.,][]{evans}.
The spectral index for the heavily absorbed
component was not well constrained, and when fixed at $\Gamma=1.7$
($\alpha_{\rm x}=0.7$) the excess absorption over the Galactic value
was characterized by an $N_{\rm H}$ of about $3 \times 10^{23}$
cm$^{-2}$. A problem remained in that the component of unabsorbed
emission remained unphysically inverted, at $\alpha_{\rm x} \approx
-1$.

We found that fits were unacceptable unless the normalization of the
highly-absorbed emission (that dominates above about 3 keV) decreased
between the 2010 and 2014 observations.
Variability was not required for the
unabsorbed component whose parameters were linked between the data
sets.  This model of a very hard, unabsorbed, low-energy component
($\alpha_{\rm x} = -1.2$) plus a component of typical (fixed) spectral
index and high absorption, gave an acceptable fit of $\chi^2=85.5$ for
70 degrees of freedom.  
Of course we cannot be sure
it is the emission that has decreased and not the absorption that has
increased, or at least some combination of the two.  But, the
alternative of keeping a fixed normalization and fitting $N_{\rm H}$
as $1.9$ and $3.6 \times 10^{23}$ cm$^{-2}$ in 2011 and 2014,
respectively, gave a somewhat worse fit of $\chi^2=91.2$ for 70
degrees of freedom, and allowing both normalization and $N_{\rm H}$ to
be free parameters found the solution where only the normalization
varied.  

\begin{figure}
\centering
\includegraphics[width=7.cm]{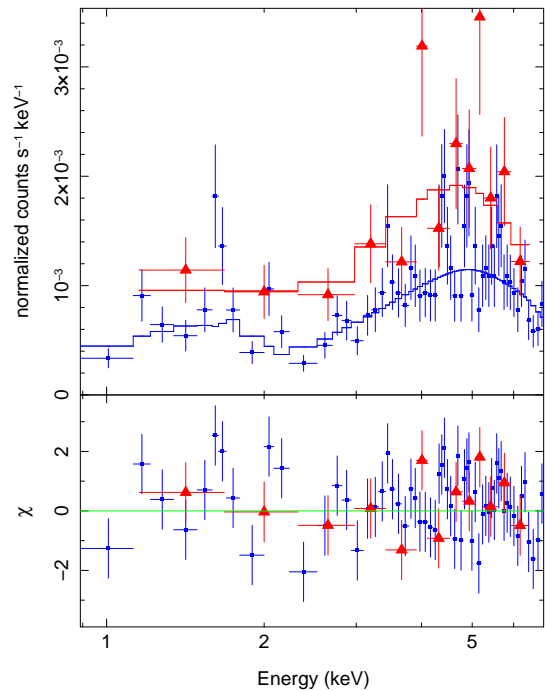}
\caption{Core spectrum fitted to a power-law model absorbed by a
  power-law distribution of column densities, as described in the
  text. The red triangles and upper model curve correspond
    to the 2010 ACIS-S data,
  and the blue squares and lower curve
  to the 2014 ACIS-I data. The lower panel shows residuals expressed as
  their individual contributions, $\chi$, to $\chi^2$.}
\label{fig:corepwab}
\end{figure}

While the model above explains some characteristics of the core
emission, it does not explain the unphysically hard spectral index
found for the unabsorbed, non-variable, emission.  We thus
investigated changing the absorption model from the simplest case of a
single column density.  A partial covering model applied to a
single-component power law of $\alpha_{\rm x} > 0$ did not give a good
fit, even if the normalization and the parameters of the absorption
were allowed to vary between 2010 and 2014 (e.g., $\chi^2=188.5$ for
69 degrees of freedom with $\alpha_{\rm x} = 0.7$).  However, a
power-law distribution of column densities ranging from $10^{15}$ to
$10^{24}$ cm$^{-2}$ (fixed), with the covering factor independent of
column density i.e., the model of \citet{donem} with $\beta=0$,
produces a marginally acceptable fit of $\chi^2=94.0$ for 72 degrees
of freedom for a fitted power-law index of $\alpha_{\rm x}=
0.69^{+0.12}_{-0.13}$, and $\chi^2$ reduces by 3 for one extra
parameter if about 0.2 per cent of the 1-keV flux density is
unabsorbed (Fig.~\ref{fig:corepwab}).  For this model the 1-keV flux
density of the absorbed emission (after correction for absorption)
dropped from about 230 nJy in 2010 to 150 nJy in 2014.  The data
quality is insufficient for us to comment reliably on the possible
presence of a 6.4-keV Fe line.  The 0.5--7~(2--10)-keV core power
(corrected for absorption) was $3 (2.5)\times 10^{43}$ and $2 (1.7)
\times 10^{43}$ ergs s$^{-1}$ in 2010 and 2014, respectively.  The
1-keV flux density in the component which is unabsorbed is only about
0.3 nJy.

While the details of the absorption are likely to be complex and
cannot be uniquely determined by these data, we conclude that the
overall core emission decreased significantly in strength between 2010
and 2014, and the material covering the source is non-uniform,
reaching a maximum column density as high as about $10^{24}$
cm$^{-2}$.  \source's 
nuclear emission-line classification is that of HEG, and
like all HEG in the low-redshift 3CR sample, 
shows morphology that is predominantly \frII\
\citep{butt2}.
Following \citet{evans}, the nuclear X-ray emission of
such galaxies is usually dominated by emission which is
absorbed by $N_{\rm H}> 10^{23}$ cm$^{-2}$ and which is supposed to
arise from structures associated with the accretion flow, with a
smaller contribution from unabsorbed emission assumed to be jet
related in origin.  \source\ appears to validate such a picture.  The
X-ray luminosity (corrected for absorption) is lower than that of a
typical \frII\ in \citet{evans}, pointing to a less luminous accretion
flow, as might be anticipated for a source of intermediate \fr\ power.

\citet{morganti} find no HI in absorption against the radio nucleus.
Given the X-ray absorption results, the logical conclusion is that the
radio jet becomes bright only after it has penetrated the environment
of the absorbing gas.  The flat radio spectrum (e.g., $\alpha_r
\approx 0.1$ between 1.4 and 5 GHz in 1981) shows it is still
moderately synchrotron self-absorbed at that point.  The geometry of
the absorbing gas is unknown, but radio-galaxy jets passing through
the holes of a toroidal structure whose axis is in the plane of the
sky is the picture generally adopted by unified models
\citep[e.g.,][]{urrypadovani} to explain the existent of beamed
sources with no central X-ray or optical absorption.  With \source, we
have perhaps the most convincing evidence in a relatively powerful
radio galaxy that the nuclear X-ray emission is dominated by a
variable sub-pc component on a scale smaller than that at which the
inner radio jets become bright.  Within the context of unified models,
the high inner absorption argues that the source is relatively close
to the plane of the sky.

Considering now the base of the radio jet, we should attribute to it
only the non-varying core X-ray component that is entirely unabsorbed
(0.3 nJy at 1 keV).  By comparison, our highest-resolution radio map
reveals the core 5-GHz flux density to have been 12.4 mJy in 1981
(Table \ref{tab:radiocomps}).  The interpolated radio to X-ray
spectral index is then $\alpha_{\rm rx} \approx 1 $ for the base of
the jet.  If we include data for the optical nucleus measured by
\citet{capetti-cores}
and interpreted as synchrotron radiation, $\alpha_{\rm ro} \approx 0.75 $
and $\alpha_{\rm ox} \approx 1.4 $.

\begin{table*}
\caption{Parameters for radio components}
\label{tab:radiocomps}
\begin{tabular}{llllllll}
\hline
(1) & (2) & (3) & (4) & (5) & (6) & (7) \\
Component & RA, Dec & $S_{\rm 1.4~GHz}$ (mJy)& $S_{\rm 4.9~GHz}$ (mJy)& FWHM (arcsec) & PA (deg)
& $\alpha_{\rm r}$\\
\hline
Core & 12:54:$11.999\pm0.007$, 27:37:$33.91\pm0.06$ & $14.2\pm0.2$ &
$12.4\pm0.2$ & -- & -- & $0.11\pm0.02$ \\
K1 & 12:54:$12.190\pm0.005$, 27:37:$28.64\pm0.05$ & $16.2\pm0.6$ &
$7.1\pm0.2$ & $0.7\times0.3$ & 21 & $0.67 \pm 0.04$ \\
K2 & 12:54:$12.174\pm0.003$, 27:37:$26.30\pm0.08$ & $26.5\pm0.6$ &
$11.6\pm0.3$& $0.6\times0.4$& 32 & $0.67\pm0.04$ \\
\hline
\end{tabular}
\medskip
\begin{minipage}{\linewidth}
$1\sigma$ errors.
\\
\end{minipage}
\end{table*}

\subsection{N hotspot}
\label{sec:nhot}%

The spectrum of the bright upstream edge of the N hotspot was extracted
from a circle of radius 2 arcsec.  Local background was measured from
a source-centered half annulus of radii 2.5 to 10 arcsec to the SE,
completely on the same CCD chip as the hotspot (I3 and S3 for the 2014
and 2011 data, respectively).  There are only about 40 net counts in
the band 0.4-7 keV, and so they were fitted using the Cash statistic
(cstat within {\sc xspec}) using a power-law model, to find a spectral
index of $\alpha_{\rm x} = 1.3^{+0.8}_{-0.7}$ and 1-keV flux density
of $0.7^{+0.4}_{-0.3}$ nJy.  Results are very similar using $\chi^2$
and Churazov weighting with 5 counts per bin.  If fitted to a thermal
(apec) model, $kT$ is hot, at about 3~keV, which does not suggest any
obvious type of physical object.  Nonthermal emission is far more
likely.

\subsection{K1 and K2}
\label{sec:k12}%

The spectral-fitting approach for K1 and K2 was similar to that for
the N hotspot, except that polygons were used.  Net counts for the
both regions agree within statistics with one another and with the N
hotspot.

Results for K2 are similar to those for the N hotspot, with a thermal
fit giving a high $kT$ (about 3.8 keV, and an abundance value that
tends to zero), while a power-law fit gives a result that is
understandable in the context of X-ray jet knots in other sources,
finding $\alpha_{\rm x} = 1.0 \pm 0.6$ with a 1-keV flux density of
$0.4^{+0.2}_{-0.1}$ nJy.  We interpret these X-rays as being
nonthermal in nature.

However, the contrary is true for K1.  A power-law spectral fit gives an abnormally steep
spectral index of $\alpha_{\rm x} = 3.2\pm 1.1$, whereas a
thermal fit gives reasonable parameter values.  The abundance is
poorly constrained at $< 1.0$ and $kT = 0.37^{+0.7}_{-0.1}$ keV.

Table \ref{tab:radiocomps} gives details of the radio emission from K1
and K2.  Similar to the inner jet/core (see above), the interpolated
radio (5 GHz) to X-ray spectral index for K2 is $\alpha_{\rm rx}
\approx 1 $.  K1 is fainter than K2 in the radio, and while we
cannot rule out some of the X-rays seen in this region as
having a nonthermal origin, the X-ray spectral data combined with the
closer morphological association of the X-rays to emission-line gas in
the jet deflection region (Section \ref{sec:hic}) than to the radio,
lead us to consider only a thermal origin for the K1 X-rays.

\subsection{Large-scale X-ray emission}
\label{sec:Xextended}%

We have used an ellipse of semi-major and semi-minor axes 28.5 and
22.3 arcsec, respectively,  enclosing the radio source,
to sample the spectrum of the large-scale gas. The background was
measured from larger source-free regions on the relevant I3 or S3
chips. Regions covering K1, K2,
the N hotspot, the core (a circle of generous radius 5 arcsec to
exclude the wings of the point spread function), and other
sources found by {\sc wavdetect} are excluded.   
The fit to a
single-temperature thermal ({\sc apec}) model is acceptable ($\chi^2 =
12.5$ for 14 degrees of freedom).
The temperature,
$kT = 1.0 \pm 0.3$ keV, is
compatible with a weak group atmosphere, although uncertainties are
large.

We checked our procedure using background measured from the 
\chandra\ blank-sky fields. We followed the method we used in
\citet{worrall-swirl}, where we found the normalization correction
needed to match source and background fields at
9.5-12-keV (where particle background dominates) was less
than 3\%.  While the same was true here for the ACIS-S data, to match
the 2014 ACIS-I data we needed to reduce the level of the
sky-background data by the
large factor of 30\%,  
perhaps as a result of the 2014
observation being made close to Solar Maximum, when particle background is reduced.
Although the large correction factor questions the appropriateness of
the sky-background files for the ACIS-I observations here,  we note that our
spectral results were in
excellent agreement with those from local background, in finding
$kT = 1.0 \pm 0.3$ keV.

The abundances found in our fits were low,
at $0.05^{+0.10}_{-0.04}$ times solar, but this may be an artifact of
fitting a simple model to a region which is likely to include
temperature structure, together with contributions from unresolved
low-mass X-ray binaries and 
X-rays arising from inverse Compton
scattering of the Cosmic Microwave Background radiation (iC-CMB) by an
electron population responsible for the radio synchrotron emission, as
detected in many lobed radio galaxies \citep[e.g.,][]{croston}.

We argue that in \source\ emission from diffuse thermal gas 
 {\it dominates\/} the 
extended X-rays.  Firstly, X-ray binaries are
estimated at only a few per cent of the
1-keV flux density based on typical integrated luminosities in
elliptical galaxies taken from \citet{kimf}.
A more important consideration is iC-CMB emission.  This
should be
strongest where the radio is brightest.  While to some degree that is
true, in that the extent of X-ray emission evident in Figure~\ref{fig:xradaptimage} is
similar to that of the overall radio source, there is a greater association with the
surrounds of the lobes than with regions where the lobes are radio
brightest (see Section \ref{sec:overall}).

For a more quantitative assessment of the iC-CMB emission we first note that
it is difficult spectrally over the \chandra\ bandpass to discriminate
between gas of temperature roughly 1 keV and steep-spectrum power-law emission
unless statistics are very good.  The difficulty applies here, and
a fit to a single-component power law is formally acceptable, but the
spectral index of
$\alpha_{\rm x} = 1.8 \pm 0.4$ should match $\alpha_{\rm r}$ for
iC-CMB, which it does not ($\alpha_{\rm r}
\approx 0.65$ to 0.68 over the brightest regions of the radio lobes).
Moreover, we have modelled the brightest parts of each radio lobe as
spheres of radii 8 arcsec, and found minimum-energy magnetic fields
(without protons) of
1.5 and 1.4 nT for the S and N lobes, respectively\footnote{Field strengths
published by \citet{vanbreugel} are slightly higher.  This is as expected since
protons
of equal energy to the electrons were used in those calculations.}.  With these field
strengths, only about 5 percent of the 1-keV flux density of extended
emission is accommodated, and the field would need to be reduced by a
factor of about 6
for iC-CMB emission to be at a level comparable to the X-ray emission
seen. 
This factor is much larger than typical of radio lobes \citep{croston}.
Even then the steep spectrum would still not be explained, and so we
consider non-thermal explanations for the extended X-ray emission no further.

\section{Discussion}
\label{sec:discussion}

\subsection{Large-scale gas}
\label{sec:gas}%

We have argued that the extended X-ray emission is dominated by
emission from gas of $kT \approx 1$~keV, a temperature appropriate for
group gas.  The inferred bolometric luminosity in a source-centre
annulus of radii 6 and 25 arcsec is $L_{\rm bol} = 3.6 \times 10^{41}$
ergs s$^{-1}$ (or $L_{\rm x} = 2.7 \times 10^{41}$ ergs s$^{-1}$ in
the commonly-used 0.1-2.4~keV band).  To test roughly if the surface
brightness lends support for a group origin, we have fitted a
$\beta$-model to a radial profile of the emission within this annulus.
The best fit gives $\beta = 0.42$ and a core radius of 7.3 arcsec,
although uncertainties are large.  Based on scaling relations of
\citet{vikhlinin} it is necessary for this system to go out to a
radius of roughly 295 arcsec to reach $r_{500}$.  If we then restrict
$\beta$ to lie in a reasonable range of 0.4--0.6, we find an
extrapolated $L_{\rm bol} (L_{\rm x})$ within $r_{500}$ of
${2.5^{+1.0}_{-1.7}} (1.9^{+0.8}_{-1.3}) \times 10^{42}$ ergs
s$^{-1}$, which is not unreasonable for a group atmosphere of this
temperature \citep*[see e.g.,][]{lovisari}.  However, it is striking
that the detected X-ray emission covers only the radio lobes.

How recently the group was formed is an interesting question.  The
NASA Extragalactic Database (NED) reveals the presence of several
fainter galaxies within a projected radius of 120 kpc (75 arcsec) of
\source.  While velocity data are largely lacking, 2MASX
J12541671+2736531, with an appearance in \hst\ data of an edge-on
spiral, lies on the perimeter of such a circle and is separated in
velocity by only about 600 km s$^{-1}$.  Thus it is reasonable to
suppose the presence of group companions.  The evidence for
\source\ expanding into at least $10^9$ M$_\odot$ of relatively
settled gas covering the lobes led \citet{morganti} to propose
that a major merger commenced at least $10^8$ years ago,
before the triggering of the radio source.  
We estimate the mass in X-ray-emitting gas
to be roughly $10^{11}$ M$_\odot$ within a radius of 25 arcsec of the
centre.  This then dominates the mass of the presumed merger remnant,
and argues that the group atmosphere is likely to pre-date the merger.

The morphological comparison between the radio and X-ray emission
(Section \ref{sec:overall}) shows evidence for the radio lobes having
bored cavities in the gas, as is found common in group and cluster
atmospheres \citep[e.g.,][]{panagoulia}.  In group atmospheres
cavities are found to be accompanied by particularly
strong shocks, with Mach numbers ranging from about 1.5 to 3, such as in HCG
62 \citep{gitti}, NGC 5813 \citep{randall} and \pks\ \citep{wfos},
although only in the last case is the radio source from the \fr\ class
of which \source\ is also a member and which is expected to be most
characteristic of heating in the Universe as a whole.  The way in
which the X-ray enhancements embrace the radio lobes of
\source\ points to relatively strong shocks being likely here too,
although the X-ray emission is too weak for quantitative assessment of
shock speed.

\citet{wfos} found for \pks\ that the 
the
cavity power, $P_{\rm cav}$, increased 
by a factor of 40 once the deduced Mach number ($\approx 3$) of the
expansion and the kinetic and thermal energy in
shocked gas are taken into account.
Without this increase, $P_{\rm cav}$ was much too low to fall on
an extension from \frI\ to \fr\ powers of  correlations
with 1.4-GHz radio power \citep{cavagnolo}.
For \source\ we cannot measure a Mach number for the expansion, but
we do find that $P_{\rm cav}$ falls far
below correlations unless there is a significant energy contribution
from strong shocks.
The total 1.4-GHz flux density from the NVSS survey
\citep{condon} is 2.9 Jy, leading to $P_{1.4} \approx 7 \times 10^{41}$ erg s$^{-1}$ and an
expected cavity power from correlations of 
$P_{\rm cav}$ of $2 \times 10^{45}$ erg s$^{-1}$, although with
large errors \citep{cavagnolo}.
In contrast, assuming the group gas in a sphere of radius $7''$ has been
pushed aside by each lobe in a lifetime of about 10 Myr (see Section
\ref{sec:hic}, a factor of
four shorter than the sound-crossing time in the group gas), with no
correction for the kinetic and thermal energy in shocked gas 
we estimate a cavity power of only $P_{\rm cav} \approx 3
\times 10^{43}$ erg s$^{-1}$.  

\begin{figure}
\centering
\includegraphics[width=8.5cm]{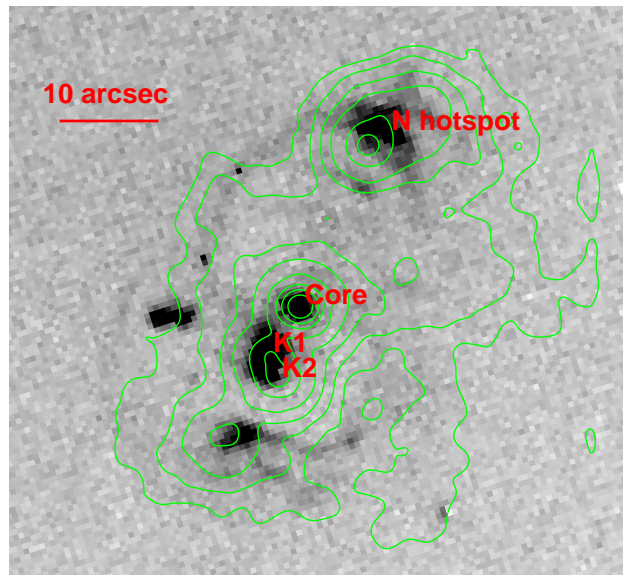}
\caption{X-ray contours (from the image of
    Figure~\ref{fig:xradaptimage}) on the H$\alpha$ image from figure
  1b of \citet{tadhunter}.}
\label{fig:largeXonHalpha}
\end{figure}

The argument that the observed X-ray gas is shocked by the radio lobes
of \source\ is also supported by a morphological comparison between
the locations of X-ray and cooler gas around the lobes.  While
the picture of enhancements embracing the radio lobes is generally
similar to that seen in H$\alpha$, an overlay of X-ray contours
on an emission-line image (Figure~\ref{fig:largeXonHalpha}) suggests
that, with the exception of the core, jet-cloud interaction region
and northern hotspot, there is an anti-correlation of X-ray and
H$\alpha$ emission, particularly for filaments of emission-line
gas in the east and south-west quadrants.  This can be understood
if shock advance around the lobe is responsible for the X-ray
brightening, and is inhibited by denser H$\alpha$-emitting material.
The anti-correlation is not easy to understand if the lobes are weakly
pushing aside the external medium, since in that case X-ray and
H$\alpha$ co-location would not be disfavoured.

The combined result of the arguments above is that the lobe expansion
of \source\ is likely to be similar to that of the
jet-cloud-interacting \fr\ source \pks.  In both cases the time
averaged jet power imparted to pre-existing group gas is primarily in
the form of kinetic and thermal energy of shocked gas.

\subsection{Jet-interaction region}
\label{sec:hic}%

\begin{figure}
\centering
\includegraphics[width=8.5cm]{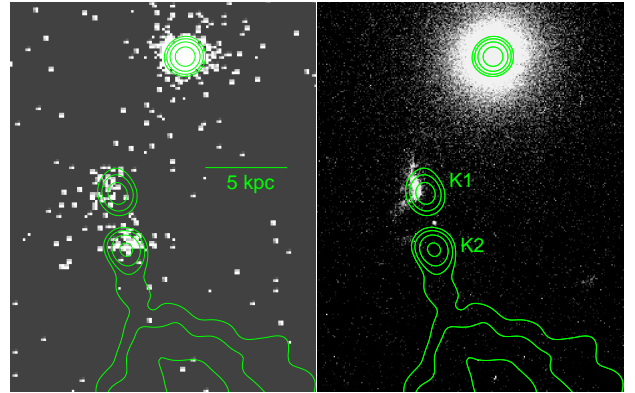}
\caption{
X-ray emission in K1 (left panel, data from Figure~\ref{fig:xrimage})
is offset relative to the radio contours in the same sense as
are the optical filaments (right panel, \hst\ WFPC2 F555W  image
formed from archival
datasets u3a12z01t and u3a12z02t).}
\label{fig:xrcorejet}
\end{figure}

\begin{figure}
\centering
\includegraphics[width=4.5cm]{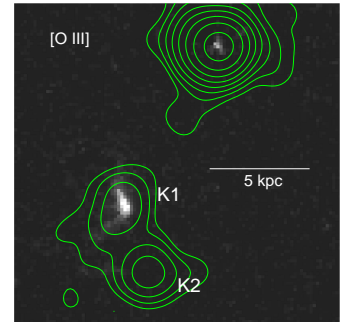}
\caption{X-ray contours (lowest of $3\sigma$ significance) on an \hst\ [O\,III] image formed from
    subtracting a scaled version of an image in F555W from that in
    FR533N. Note the X-ray and [O\,III] align well at K1 and the nucleus.}
\label{fig:OIII}
\end{figure}

The eastern displacement with respect to the radio of the X-rays
associated with K1 is what is expected if we are seeing hot gas
associated with the cloud of previously-known, cooler, optical emission-line gas.
The emission-line cloud was first reported by \citet{miley}.  It was
explored further by \citet{vanbreugel} who
found a
mass\footnote{scaled up by a factor of two following note
  added in proof.} of  about $4 \times
10^6$ M$_\odot$, and
argued on energetic grounds
that it was local gas that is being heated and entrained by the jet
rather than having been transported from the nucleus.   The X-ray
detection makes that conclusion even more solid, since the cloud mass
grows by more than an order of
magnitude once the mass of the X-ray-emitting envelope is taken into
account (see below).  The required spatial resolution for improved
mapping of the morphology of the optical gas became available with \hst, and \citet{martel}
commented on the cloud's north-south orientation and filamentary appearance
based on 1995 observations using the F702W filter [see also \citet{capetti}
who incorporated F555 data].  In
Figure~\ref{fig:xrcorejet} we show archival \hst\  continuum data from 1997
alongside an X-ray image. In Figure~\ref{fig:OIII}, X-ray
contours are overlaid on a continuum-subtracted narrow-band \hst\ image
centered on the [O\,III] emission line.  Despite the poorer X-ray
resolution,
the association between the X-ray and
optical emission is clear.  These emissions are adjacent to,
rather than coincident with, the radio peak at K1.

We estimate the physical properties of the X-ray gas associated with
the K1 gas cloud assuming the emission fills a sphere of radius 1.1 arcsec (diameter
3.5 kpc) with uniform density, using equations from section 4.4 of
\citet{wfos}.  We find a density of hydrogen nuclei $n_{\rm p} =
2^{+0.5}_{-1.5} \times 10^5$ m$^{-3}$, a pressure of $P =
3^{+4}_{-1.6} \times 10^{-11}$ Pa, and a total X-ray gas mass of
$M_{\rm Xgas} = 17^{+4}_{-13} \times 10^7$ M$_\odot$.   If the X-ray gas is not still
heated, e.g., by the radio jet, the cooling time is $60^{+120}_{-20}$
Myr.  
The average pressure in the group X-ray gas is found to be roughly
10 and 4 $\times 10^{-13}$~Pa at radii of 10 and 25 arcsec,
respectively.
At a radius of 6 arcsec, comparable to that of the K1 gas cloud
we find $1.5^{+0.3}_{-0.4} \times 10^{-12}$ Pa.
The group gas cannot therefore
contain the gas cloud, which should expand and dissipate in the
sound crossing time of about 10 Myr. We infer that the cloud was originally cooler
material that has been heated to X-ray temperatures within the
lifetime of the radio source due to the passage of the jet.

We note that the X-ray properties of the gas cloud are not dissimilar
to those found in the closer, and so brighter and better constrained,
high ionization cloud (HIC) interacting with the jet in \pks.  Indeed,
for \pks's HIC, $P \approx 8 \times 10^{-11}$ Pa, $M_{\rm Xgas}
\approx 7 \times 10^7$ M$_\odot$, $kT \approx 0.3$ keV, and the gas cloud
was heated within the lifetime of the radio source \citep{wfos}.
Differently from \pks, in the HIC of \source\ the optical emission
lines are narrow and in a high ionization state, with a total flux in
[O\,III] of $3 \times 10^{-14}$ ergs cm$^{-2}$ s$^{-1}$ \citep{si}.
The narrowness of the lines was unexpected because the 
jet-cloud interaction appears to be strong, but one of the possible
explanations proffered by \citet{si} is supported by our detection of
X-rays associated with the cloud.  The idea is that the optical
emission-line gas is not itself shocked, but lies around the shocked
structures, and is photoionized by photons emitted by gas behind the shock front.
The [O\,III] line luminosity is high, at $5 \times 10^{41}$ ergs
s$^{-1}$, but the X-ray luminosity (extrapolated from the fitted parameter values
to include all photons above 13.6~eV), at
$2.4^{+6.5}_{-1.5} \times 10^{41}$ ergs
s$^{-1}$, can provide a significant fraction of the
required photoionizing power. The gas that is shocked by interaction with the
radio jet is likely to be structurally complex, and lower-temperature
components (e.g., in the ultraviolet) could plausibly make up any
deficit.

The isotropic X-ray
luminosity of the nucleus is about $2 \times 10^{43}$ ergs s$^{-1}$,
after correction for the high absorption
(Section \ref{sec:core}).  If the nuclear X-ray spectrum extends to high
energy without downwards curvature, this would be an underestimate. 
The cloud could intercept about 1 per cent of this luminosity, and so
the contribution of ionizing photons from the nucleus will be similar
to, or could exceed, the contribution from the X-ray-bright gas at K1.
We note that \citet{ramirez} contradicts the high polarization at K1
reported by \citet{miley}, and so there is no evidence that the
optical continuum at K1 arises from scattered nuclear light.

We have found above that the X-ray gas cloud can be no older than 10
Myr.  It is then interesting to attempt to relate this to the likely
age of the radio source.  From Figure~\ref{fig:radio} we see that the
southern lobe is morphologically consistent with plasma originating
from a jet travelling from K1 to K2, rather than from an historically
un-deflected jet.  Although we note that both lobes show more
distortion on the east than west, the side on which more of the
H$\alpha$-emitting filaments mapped by \citet{tadhunter} lie, it seems
more plausible that the southern jet was bent early on in the lifetime
of the source than being a late-phase re-direction.  This makes
\source\ no older than about 10 Myr.  Such an age fits with the
source's relatively small projected linear size and our argument
(unlike for \pks\ which is of similar projected size) that it lies
close to the plane of the sky, and so is of relatively small intrinsic
size.

The nature of the jet-cloud interaction will be explored further using
new Integral Field Unit data from Gemini North (Duncan Smith et al,
work in progress).

\subsection{Nonthermal X-ray components}
\label{sec:Xnonthermal}%

The X-ray emission associated with K2 peaks upstream of the radio.
This has been seen elsewhere in X-ray synchrotron knots embedded in the
jets of nearby radio galaxies, most notably in 3C\,66B, Cen\,A,
3C\,346, 3C\,353 and
 possibly 3C\,15
\citep*{hard-66b, hard-cena,
worr-3c346, kataoka, dulwich-3c15}.  The spectral constraint on the X-ray emission from
knot K2 is rather poor and alone does not rule out an origin as either
inverse Compton emission from electrons of energies similar to those
producing the radio synchrotron emission, or synchrotron radiation
from high-energy electrons of steeper spectrum.  The deflection at K1
has been interpreted by \citet{tingay} in the context of an
oblique-shock model to support a pre-shock jet Lorentz factor of
between 3 and 5, a post-shock Lorentz factor of between 1.1 and 1.5,
and a jet orientation of less than $75$ degrees to the line of
sight. Even if the post-shock orientation angle is as small as a few
degrees (unlikely due to the X-ray absorption against the nucleus,
Section \ref{sec:core}), the Doppler factor in K2 will be insufficient to
produce the X-ray emission through boosting of inverse-Compton
scattered photons, as may be important in quasar jets \citep[see e.g.,
][and references therein]{wrev, marshall}.  
For negligible beaming, predictions for synchrotron
self-Compton emission and Compton scattering of CMB photons, based on
a minimum-energy magnetic field, fall short of observations by factors
of roughly 1300 and 600, respectively. However, K2 is only 13 kpc from
the nucleus, and starlight will provide additional photons. To
estimate their contribution we have assumed the host galaxy to have
similar properties to that for \pks, for which we modelled the local
energy density of starlight as a function of position \citep{wfos}.
We find an X-ray contribution that is roughly five times the
contribution from unscattered CMB radiation -- still far below what is
required.

We therefore conclude that the X-ray emission in K2 is dominated by
synchrotron radiation.  This is the established mechanism for
X-rays from \frI\ radio galaxies \citep*{worr-fr1s, harwood}.
It was first claimed in knots of specific \frII\ radio
galaxies by \citet*{wilson-pica, worr-3c346, kraft-3c403, kataoka}, 
and is studied in detail in the bright jet of
Pictor A \citep{hard-pica}. X-ray synchrotron jet emission is
also seen in the \fr\ transition source \pks\ \citep{wfos}, where as
in \source, $\alpha_{\rm rx} \approx 1$.

Bright nonthermal X-rays on the upstream edge of a radio hotspot 
(Sections \ref{sec:overall} and \ref{sec:nhot})
have been noted
previously in other sources \citep*[e.g.,][]{erlund, hard-hot}, with a variety
of suggested explanations. Indeed an offset is seen in
\pks, but with the main difference that the offset in that source is for the
hotspot on the jet-side of the source, approaching at small angle to the line of
sight.  Following \citet{georg}, \citet{wfos} argue
that inverse-Compton scattering of hotspot synchrotron emission
by the approaching jet may provide an explanation for that offset.
\source\ has only one (N) radio hotspot, but as it is
on the side of the source highly unlikely to be approaching the observer,
a different explanation must apply.

At 0.4-arcsec resolution the radio map of the N hotspot starts to
break into faint radio substructure. \citet{tingay-pica} have
suggested, based on VLBI imaging principally of Pictor A, that
variable pc-scale regions host strong shocks that accelerate electrons
to TeV energies and give significant X-ray synchrotron emission. This
may be happening in \source.  In a minimum-energy magnetic field it
would take the whole flux density of the hotspot (about 113
mJy at 4.9 GHz) in a pc-scale region to produce the X-rays via the
synchrotron self-Compton mechanism.  Instead, overlooking the
non-contemporaneous nature of the radio and X-ray data, at most 3~mJy
of radio flux density comes from the X-ray-bright region, but this
implies $\alpha_{\rm rx} \approx 0.9$, which is similar to the value
of $\alpha_{\rm rx}$ in K2 for which we have argued a synchrotron
origin.  Unfortunately the hotspot X-rays are too faint for us to
comment on possible variability between the 2010 and 2014 data, which
might address the issue of source size, but more sensitive
high-resolution radio mapping closer in date to the X-ray observations
could help explore the likely synchrotron origin further.

\section{Summary and Conclusions}

The radio galaxy \source\ is an ideal test-bed for studying the energy
deposition of a radio galaxy typical of those dominating radio-mode
feedback in the wider Universe, since it has typical
power and resides in a typical (group) environment.
Remnants of cold and excited gas from the merger that probably
triggered the radio outburst have previously been studied dynamically
through their line emission.
In this paper we have reported the discovery with \chandra\ of
X-ray-emitting gas dominating the baryonic mass of the system both on the group
scale and in the jet-cloud interaction region that is responsible for
having bent the jet by a projected angle of $40^\circ$.  Our primary
conclusions can be summarized as follows:

\begin{enumerate}

\item
The expanding radio galaxy has been responsible for shocking a
pre-existing X-ray group atmosphere which now, to some extent,
embraces the lobes. We estimate the mass of group gas out to a radius
of 40 kpc as $10^{11}$ M$_\odot$.
Shock advance in some places is inhibited by the cold merger material.
The kinetic and thermal energy of shocked gas then dominates the overall heat
input from the radio galaxy to the external medium.

\item
The emission-line cloud responsible for bending the jet is dominated
by X-ray gas of mass about $2 \times 10^{8}$ M$_\odot$.
The X-ray spectrum of the nucleus contains a highly absorbed component
arguing that the galaxy lies close to the plane of the sky.  The large
projected jet bend is then also a large intrinsic bend.

\item
The emission-line cloud can be no older than 10 Myr or it should have
expanded away.  The fact that the radio lobe
(containing old jet material)  is
inflated around the post-bend jet direction allows us to date the
whole radio source as no older than 10 Myr.

\item 
The discovery of dominant X-ray-radiating gas associated with the
emission-line cloud goes some way to solving the mystery of the
narrowness of the optical emission lines in the cloud.
The X-ray material can provide about  $2 \times 10^{41}$ ergs s$^{-1}$
to photoionize the cloud, suggesting the optical emission-line gas may be
unshocked.  Ionizing continuum from the core could provide a similar
level of excitation at the cloud.

\item
X-ray emission is detected from the southern jet and northern hotspot,
in both cases offset upstream from radio peaks.  We argue that in both
cases the X-ray emission is likely to be synchrotron radiation.

\end{enumerate}

The closer source \pks\ shares many of the properties of \source, such
as typical radio power, shocked group atmosphere, and is also further
enriched by revealing a jet-cloud interaction region.  As compared
with \source, the radio outburst of \pks\ is estimated to be somewhat
older, there is no longer evidence of large-scale merger gas (if
indeed it existed), and the jet-cloud interaction may be younger and
more localized, with the jet having multiply scarred the cloud with
which it is interacting.  Both sources have grown into the plateau
stage where they are expected to spend most of their lives, and so can
be considered as touchstones for revealing how typical radio-mode
heating may occur in the Universe.

\section*{Acknowledgments}

We thank Clive Tadhunter for providing the H$\alpha$ data from
\citet{tadhunter}
that we use in Figure \ref{fig:largeXonHalpha}.
We acknowledge support from NASA grant GO3-14118X.  Results are
largely based on observations with \chandra, supported by the CXC.
The National Radio Astronomy Observatory is a facility of the National
Science Foundation operated under cooperative agreement by Associated
Universities, Inc. 
This research has made use of the NASA/IPAC Extragalactic Database
(NED) which is operated by the Jet Propulsion Laboratory, California
Institute of Technology, under contract with the National Aeronautics
and Space Administration. 


\end{document}